\newcommand{\be}{\begin{equation}}
\newcommand{\ee}{\end{equation}}

\documentstyle[aps,epsf,multicol]{revtex}
\begin{document}
\draft
\title{The $\theta$ vacuum reveals itself as the fundamental theory of the 
quantum Hall effect ~~~~~~~~~~~~I. Confronting controversies}
\author{A.M.M. Pruisken$^{1,2}$, M.A. Baranov$^{2,3}$ and M. Voropaev$^{2}$ }

\address{$^{1}$ Indian Institute of Science, Bangalore 560012, India}
\address{$^{2}$ University of Amsterdam, Valckenierstraat 65, 
1018XE Amsterdam, 
The Netherlands}
\address{$^{3}$ RRC "Kurchatov Institute", Kuchatov Square 1, 123182 
Moscow, Russia}

\maketitle

\begin{abstract}
\noindent{The $\theta$ dependence of the Grassmannian 
$U(m+n)/U(m)\times U(n)$ non-linear $\sigma$ model is reexamined. 
This general theory provides an important laboratory for studying the 
quantum Hall effect, in the special limit $m=n=0$ (replica limit). 
We discover however that the quantum Hall effect is in fact independent of 
this limit and exists as a generic topological feature of the theory for all 
non-negative values of $m$ and $n$.}
The results are in concflict with many of the historical ideas and expectations on the basis 
of the large $N$ expansion of the $CP^{N-1}$ or
$SU(N)/U(N-1)$ model 
\end{abstract}


\begin{multicols}{2}


The quantum Hall effect (qHe) is one of the richest and most interesting 
realizations of the topological concept of an {\em instanton vacuum}
\cite{llp,ammp}. Levine, Libby and Pruisken
originally elaborated on the dramatic consequences of having a $\theta$ 
{\em term} in the {\em replica field theory} representation of Anderson 
localization, by capitalizing on the existing
similarities between the theory and QCD. Along with the theory came
the promissing idea that the qHe can be used as a laboratory where the 
strong coupling problematics of quantum field theory can be studied and 
investigated in detail. This is in complete contrast to QCD and other  
theories where experiments are hardly possible. 

Striking progress has been made over the years, both from the experimental\cite{expt} 
and theoretical\cite{psb} side. 
What has emerged from the experiments on the qHe are several novel features of 
$\theta$ dependence in asymptotically free field theory
that were previously unrecognized such as $\theta$ {\em renormalization}~\cite{bf},
 a {\em massless phase} at $\theta=\pi$ \cite{ammp} etc.

However, for a variety of reasons it has remained unclear whether the 
qHe teaches us something {\em fundamental} about the instanton vacuum, 
or whether it is an interesting but otherwise highly special application 
of topological ideas in Condensed Matter Theory. 
In order to pursue the physical objectives of the qHe it 
is necessary to understand the mathematical peculiarities of the 
theory in the so-called {\em replica limit}. 
There are, however, well known and related examples of the $\theta$ vacuum 
concept, such as the large $N$ expansion of the 
$CP^{N-1}$ model~\cite{cole,dadda,wit,aff}, that are
in direct conflict with the basic features of the qHe. 
Historically of interest as an `exactly solvable' laboratory 
of QCD, the large $N$ expansion has mainly produced 
an ``arena of bloody controversies''~\cite{cole} on fundamental issues like the 
{\em quantization of topological charge}, the meaning of {\em instantons} and 
{\em instanton gases} etc~\cite{wit,aff}.
Besides all these complications the most fundamental issue, the {\em robustness} 
and {\em precision} of the qHe, has remained unexplained.

The main objective of this Letter is to address these controversies
and to provide a physical clarity that has so far 
been lacking. 
This is done by elaborating on the subtleties of a previously overlooked, new
ingredient of the instanton vacuum concept, the 
{\em massless chiral edge excitations}~\cite{psb}. This has major consequences 
for a general understanding of the theory on the strong coupling side. 
Our investigations apply to a class of topologically equivalent non-linear 
$\sigma$ models in two dimensions, defined on the Grassmann manifold 
$U(n+m) / U(n)\times U(m)$. It contains the electron gas ($n=m=0$), 
the $O(3)$ model ($n=m=1$) as well as the $CP^{N-1}$ model ($n=N-1,m=1$) as 
well known examples. Our results indicate that the fundamental features of 
the qHe are essentially topological in nature and displayed by all (non-negative)
members $m$ and $n$ of the Grassmann manifold. These features are

\begin{enumerate}
\item
\noindent
There is a {\em critical phase} at $\theta = (2k+1) \pi$ where the system has 
{\em gapless} excitations.
\item
\noindent
Away from criticality the system has always a {\em massgap} in the bulk. 
However, for $|\theta| >\pi $ there are {\em massless chiral edge excitations}. 
\item
\noindent
The theory displays {\em robust} topological quantum numbers that explain the 
{\em stability} and {\em precision} of the quantum Hall plateaus. Adjacent 
{\em quantum Hall phases}, labelled by integers $k$ and $k+1$, are seperated 
by a {\em continuous transition} that occurs at $\theta =(2k+1)\pi$. 
\end{enumerate}

\noindent{The theory} depends on $m,n$ 
only as far as the {\em details} of quantum criticality at $\theta =(2k+1)\pi$
are concerned such as the numerical value of the critical indices that enter 
into the scaling functions of the quantum Hall plateau {\em transitions}. 
The quantum critical details, however, should not be confused with the 
{\em universal topological features} of the theory (1-3) which are independent 
of $m,n$ and this includes the mathematical peculiarities that are associated 
with the {\em replica limit} ($n=m=0$).

In this Letter we use the large $N$ expansion to examine the principal features 
of $\theta$ dependence that historically remained undiscovered~\cite{pbv}. These include 't Hooft's 
idea of using {\em twisted boundary conditions}\cite{llp} as a probe for a 
massless phase at $\theta=\pi$ as well as the phenomenon of 
{\em broad conductance distributions} at quantum critical points~\cite{cp}.

\noindent{$\bullet$} {\em The action.~~~}
The Grassmannian non-linear $\sigma$ model 
involves unitary matrix field variables $Q$ that
obey the non-linear constraint $Q^2 = {\bf 1}_{m+n}$.
The action is given by     
\be
	S_{eff} [Q]  =  S_\sigma [Q] + 2\pi i \sigma_{xy}^0 C[Q],
\ee
where $S_{\sigma}$ denotes the ordinary non-linear sigma model
and $C[Q]$ the {\em topological charge}\cite{llp}  
\begin{eqnarray} 
	S_{\sigma} & = & \frac{\sigma_{xx}^0}{8} 
	\int d^2 x \, {\rm tr} \nabla_i Q \nabla_i Q
 	+\omega \rho_{bulk} \int d^2 x {\rm tr} \Lambda Q  \\
	C[Q] & = & \frac{1}{16\pi i} \int d^2 x \, {\rm tr} \epsilon_{ij}  
	Q \nabla_i Q \nabla_j Q .
\end{eqnarray}
Here, $\sigma_{xx}^0 , \sigma_{xy}^0$ represent the dimensionless 
longitudinal and Hall conductivities 
in mean field theory respectivily, the $\rho_{bulk}$ equals the density of 
electronic levels and $\omega$ is the external frequency. 
By using the representation 
\begin{eqnarray}
  Q=T^{-1} \Lambda T ;~~~T \in U(m+n)~~~
	\Lambda =
  \left[\matrix{{\bf 1}_m & 0 \cr 0 & -{\bf 1}_n }\right] ,
\end{eqnarray}
one can express $C[Q]$ as an integral over the edge of the system

\be
	C[Q] =  \frac{1}{4\pi i} \oint d x \, {\rm tr} T \nabla_x T^{-1} \Lambda  .
\ee
Following the formal homotopy theory result $\pi_2 (G/H) = \pi_1 (H) = Z$ 
the topological charge $C[Q]$ is {\em integer} quantized provided $Q$ 
reduces to a constant  $\Lambda$ at spatial infinity or, equivalently,
$T$ equals an arbitrary $U(m) \times U(n)$ gauge at the edge of the system. 
\noindent{$\bullet$ }{\em Action for the edge.~~~}
As an extremely important physical limit of the theory we next address the 
situation where the Fermi energy of the disordered Landau level system is 
located in an {\em energy gap} or {\em Landau gap} \cite{psb}. 
As is well known, the low energy dynamics of the electron gas is now solely 
determined by the {\em massless chiral edge states}. 
To understand what it means in replica field theory we put
$\rho_{bulk}=\sigma_{xx}^0 = 0$ whereas $\sigma_{xy}^0$, which generally
is given as the filling fraction $\nu$ of the Landau level system, now equals 
the number $k$ of completely filled Landau levels, $\sigma_{xy}^0 = \nu = k$. 
We make use of the simplicity of the theory and split the matrix field variables 
$Q=t^{-1} Q_0 t$ into {\em edge} components $t$ and {\em bulk} components
$Q_0$. The matrix field variable $Q_0$ obey {\em spherical boundary conditions} 
(i.e. $Q_0 = \Lambda$ at the edge) such that $C[Q_0]$ is integer quantized.
On the other hand, $t \in U(n+m)$  represent 
the fluctuations about spherical boundary conditions,
$-\frac{1}{2} < C[t^{-1}\Lambda t] < \frac{1}{2}$. The $C[Q]$ can then be 
written as the sum
\be
C[Q] = C[t^{-1} Q_0 t] = C[Q_0 ] + C[ q ] ; ~~~q=t^{-1} \Lambda t ,
\ee
and the complete action becomes
\be
S_{eff} [q] =
\oint d x \, {\rm tr}\left( \frac{k}{2} t \nabla_x t^{-1} \Lambda  
+ \omega \rho_{edge} \Lambda q \right).
\label{seff}
\ee
The action (\ref{seff}) depends on the {\em edge} field variable $q$ only.
The part with $C[Q_0]$ gives rise to unimportant phase factors and 
is dropped. Next, we have added a term proportional to $\rho_{edge}$ 
indicating that although the Fermi energy is located in a Landau gap 
($\rho_{bulk} =0$), there still exists a finite density
of one dimensional {\em edge states} that can carry the Hall current.

The one dimensional theory Eq. (\ref{seff}) describes 
{\em massless chiral edge excitations} and can be solved exactly. 
Some important correlation functions of the edge theory are \cite{psb}
\be
<q> = \Lambda,
\ee
indicating that the $U(n+m)$ symmetry is permantently broken at the edge of the
system. Next
\be
<q_{+-}^{\alpha\beta} (x) q_{-+}^{\beta\alpha} (x') >  =\frac{4}{k} 
\Theta(x'-x) e^{ -\frac{4\rho_{edge} \omega}{k}(x'-x)}
\ee
Here, $q_{+-}^{\alpha \beta }$ represents an element of the $n\times m$ 
off-diagonal block of $q$ and $\Theta(x'-x)$ is the step function. 
The integer $k$ equals the number of different edge modes and the quantity 
$\rho_{edge} /k$ can be identified as the {\em drift velocity} of the chiral 
edge electrons. The most important feature of these results is that they are 
{\em independent} of the number of field components $m$ and $n$. 

The action $S_{eff} [q] $ has a quite general physical significance. It is the 
{\em replica-field-theory-representation} of the {\em integral quantum Hall state}
and is completely equivalent to the phenomenological theory of {\em chiral edge 
bosons}. However, unlike the latter approach which only applies to the
special set of integer filling fractions  $\nu = k$, the present theory can be 
extended to include the most fundamental aspect, the generation of a 
 {\em massgap} in the system rather than an {\em energy gap}. As will be shown 
next, Eq. (\ref{seff}) generally appears as the {\em fixed point action} of the strong 
coupling phase, describing the low energy dynamics of the system in the presence 
of disorder with the integer $k$ playing the role of the {\em quantized Hall 
conductance}. 
At the same time, since these conclusions can be drawn independently of $m$ 
and $n$, the chiral edge theory of Eq. (\ref{seff}) provides the resolution of the 
strong coupling problematics of the instanton vacuum concept in quantum 
field theory.

\begin{figure}
\epsfxsize=200pt
\epsfysize = 200pt
\epsffile{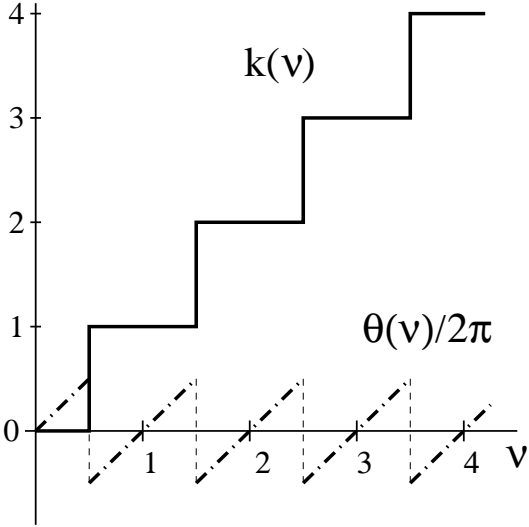}
\caption{The {\em quantized} edge part $k(\nu)$ and {\em unquantized} bulk part $\theta(\nu)$
(see Eq. \ref{C}).}
\label{Fig1}
\end{figure}

\noindent{$\bullet $}{\em The background field method.~~~}
To address the general problem of {\em mass generation} we are guided by the 
microscopic origins of the action. In particular we can write
$\sigma_{xy}^0 = \nu = k(\nu) + {\theta(\nu)}/{2\pi}$ 
where $k(\nu)$ is an exactly quantized {\em  edge} piece and $\theta(\nu)$ an 
unquantized {\em bulk} piece $-\pi <\theta (\nu )<\pi$ (Fig. \ref{Fig1}). 
The topological piece of the action becomes
\be
2\pi i \sigma_{xy}^0 C[t^{-1} Q_0 t] = i \theta (\nu) C[t^{-1} Q_0 t] 
+ 2\pi i k(\nu) C[q].
\label{C}
\ee

Notice that the quantized {\em edge} piece $k(\nu)$ is 
completely decoupled from the {\em bulk modes} $Q_0$. It gives rise to 
phase factors that can be dropped. Unlike $k(\nu)$, however,
the {\em bulk} quantity $\theta(\nu)$ generally depends on the renormalization 
of the {\em bulk} field variables $Q_0$ in a non-perturbative manner 
($\theta$ {\em renormalization}). 
To discuss this renormalization we introduce an effective action
for the edge matrix field variable $q$

\begin{eqnarray}
	S_{eff} [q] & = &  S'_{bulk} [q] + 2\pi i k(\nu) C[q]  
\nonumber \\
e^{ S'_{bulk} [q]} & = & \int_{\partial V} D[Q_0 ] 
e^{ {S}_{\sigma}^0 [t^{-1} Q_0 t] 
+  i \theta (\nu) C[t^{-1} Q_0 t]  }.
\label{Sb}
\end{eqnarray}
The action ${S}_{\sigma}^0$ has the same meaning as $S_\sigma$ except that 
the frequency term $\omega \rho_{bulk} {\rm tr} \Lambda Q$ has been dropped in 
favor of the finite system size $L$ that we shall use to regulate the 
infrared of the theory.
The subscript $\partial V$ indicates that the 
integral is performed with a fixed value $Q_0 = \Lambda$ at the edge. 

The definition of $S'_{bulk} [q]$ is precisely the same as in the 
{\em background field} methodology that has been introduced previously
for the purpose of defining the renormalization of the theory\cite{bf}. 
This methodology is based on the fact that
$S'_{bulk}$ has the same form as the original action

\be
	{S'_{bulk} [q]} = F(\theta ) + { S'}_{\sigma}  [q] + 2\pi \theta' C[q],
\label{Sb2}
\ee
except that the parameters $\sigma_{xx}^0$ and $\theta(\nu)$ are now 
replaced by {\em renormalized} ones, ${\sigma'_{xx}}$ and $\theta'$
respectivily, defined for lengthscale $L$. The free energy $F(\theta)$ is
periodic in $\theta (\nu)$.

It is easy to understand why the background field methodology plays a 
fundamental role in this problem. 
Notice that the renormalized theory measures, by definition, the sensitivity
of the {\em bulk} of the system to a change in the boundary conditions. 
Consequently, if a {\em massgap} is generated in the {\em bulk} of the sample 
then for large enough system sizes
$L$ the quantities ${\sigma'_{xx}}$ and $\theta'$ must both be zero, i.e. the
system is {\em insensitive} to changes in the boundary conditions (except for
corrections that are exponentially small in $L$). Under these circumstances,
$S_{eff} [q]$ reduces to the earlier discussed action for 
{\em massless chiral edge excitations} whereas the integer $k(\nu)$ can now be 
identified as the {\em quantized Hall conductance}. 

Since the theory is asymptotically free in two dimensions, for all non-negative 
$m$ and $n$, one expects the qHe to be a {\em generic} strong coupling
feature of the instanton vacuum, independently of the value of $m$ and $n$. 
Recall that these general statements are completely supported by the extended
weak coupling analyses (both perturtive and non-perturbative, 
i.e. instanton calculus\cite{bf}) which provide most of the phase structure of 
the theory. What has been lacking so far is an explicit example where 
these novel strong coupling features can be explored and investigated 
in detail.

\noindent{$\bullet $}{\em The $CP^{N-1}$ model.~~~}For this purpose, we next 
specify to the case where $m=1, n=N-1$ in $1+1$ dimension. 
By introducing an $N$ component complex vector field $\vec{z}$

\be 
Q_{ab} = 2z_a^*z_b -\delta_{a,b} ;~~~\vec{z}^* \cdot \vec{z} =1 ,
\ee
one can rewrite the action as a $U(1)$ gauge theory \cite{wit} 

\begin{eqnarray}
	S_{bulk} [Q] & \rightarrow & S_\sigma^0 [\vec{z} ,A_\mu ]  
	+  i \theta (\nu ) C[A_\mu ] ;
	\label{A} \\
	S_\sigma^0 [\vec{z} , A_\mu ] & = & \sigma_{xx}^0 
	\int  | D_\mu z_a |^2 ;~~~
	C[A_\mu ] =  \frac{1}{2\pi} \int \epsilon_{\mu\nu} \partial_\mu A_\nu  ,
\nonumber 
\end{eqnarray}
where $D_\mu = \partial_\mu +iA_\mu$. Within the large $N$ expansion one 
eliminates the vector field in a standard manner, and Lorentz invariance 
implies that the effective action for the $A_\mu$ field can be 
written as \cite{cole,dadda,wit,aff}

\be
	S_\sigma^0 [A_\mu ] = \frac{N}{48\pi M^2} \int d^2 x F_{\mu\nu}^2 .
\label{S0}
\ee
Here, $M=\mu_0 \exp(-2\pi \sigma_{xx}^0 /N)$ is the dynamically generated mass of the 
$\vec{z}$ vector field and $\mu_0$ an arbitrary momentum scale.

Next we emphasize that the subtleties of chiral edge dynamics
fundamentally alter many of the historical ideas\cite{wit,aff} that were
obtained on the basis of Eqs. (\ref{A}) and (\ref{S0}) alone \cite{pbv}. 
To demonstrate this
we must separate the integral pieces of the topological charge from the 
unquantized piece. The simplest way to proceed is by enforcing the constraint
$C[A_\mu ] =n + C[q]$
upon the $A_\mu$ integrals. This is done by introducing an auxiliary variable 
$\lambda$
\begin{eqnarray}
& & 	S_{bulk} [t^{-1} Q_0 t]  \rightarrow  
S_{bulk} [A_\mu , \lambda , n , q] =
\nonumber \\
	& & ~~~~~~S_\sigma^0 (A_\mu )  +  i (\theta (\nu ) + \lambda') C[A_\mu ]
	+ i\lambda' (n + C[q]).
\end{eqnarray}
By eliminating the free $A_\mu$ and $\lambda$ fields one obtains two equivalent 
expressions for $S'_{bulk}$ 
($L^2$ denotes the space-time area)  

\begin{eqnarray}
	S'_{bulk} [q] & = & -\ln \sum_{n} 
	e^{ -\frac{\pi N}{12 M^2 L^2} (n+C[q])^2  
	-i \theta (\nu ) (n+C[q])} , \\
	& = & -\ln \sum_l e^{  - \frac{3 M^2 L^2 }{\pi N} (\theta(\nu) + 2\pi l)^2 
+2\pi i l C[q]} .
\label{Sbulk}
\end{eqnarray}
The second series is obtained making use of the 
Poisson formula $\sum_{n}\exp(i\lambda n) 
=  2 \pi \sum_{l} \delta (\lambda - 2\pi l)$.
We are interested in the asymptotics $L\rightarrow \infty$
for a fixed large value of $N$. Expanding Eq. (\ref{Sbulk}) in powers of $C[q]$ we can
write

\be
S'_{bulk} [q] = F(\theta) + i{\theta}' C[q] + ...
\label{F}
\ee
where $F(\theta)$ is periodic in $\theta(\nu)$ as it should be and
\be
\theta' =  \pm 2\pi \frac{ e^{-L^2 /\xi^2 (\nu)}}{1+e^{-L^2/\xi^2
(\nu)}} ;~~~\xi(\nu) = \frac{\xi_N }{\sqrt{\pi - |\theta(\nu)|}} .
\label{Theta}
\ee
The $\pm$ sign holds for positive and negative $\theta(\nu)$ respectivily
and $\xi_N = \sqrt{N/12M^2}$. By comparing Eqs. (\ref{F}) and (\ref{Theta}) with 
Eqs. (\ref{Sb}) and (\ref{Sb2}) 
one readily concludes that the large $N$ theory displays all the aforementioned 
strong coupling features of the qHe. There is a continuously diverging
correlation length $\xi$ with a critical exponent $1/2$ which means that the 
phase at $\theta (\nu )=\pi$ is $gapless$. Away from the singular point
the Hall conductance is quantized with corrections $O(e^{-L^2 /\xi ^2})$.

\begin{figure}
\epsfxsize=200pt
\epsfysize=200pt
\epsffile{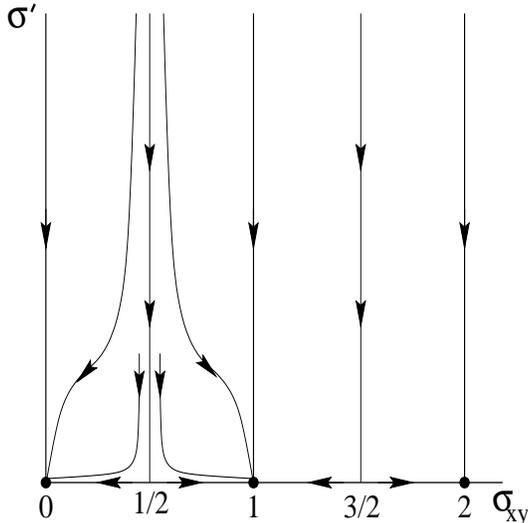}
\caption{Large N renormalization group flow diagram for 
$\sigma' =\sigma'_{xx} /N$
and the ``Hall conductance'' $\sigma'_{xy} = k(\nu) + \theta'/2\pi$. 
The arrows indicate the direction toward the infrared.}
\label{Fig2}
\end{figure}

Notice that the part $S'_\sigma [q]$ in Eq. (\ref{F}) is missing. The reason is that 
theory has been evaluated in a regime where 
$\sigma'_{xx} \approx N\exp(-ML) << N/ \exp \sqrt{\pi N/3}$ is effectively
zero~\cite{pbv}. Next, introducing $\sigma' = \sigma'_{xx} /N$ we obtain the 
asymptotic strong coupling results in differential form 
\be
\frac{d \ln \sigma'}{d\ln L}  = \ln \sigma' ,~~
\frac{d \ln |\theta'|}{d\ ln L} = 
\bigl[
2- |\frac{\theta'}{\pi}|
\bigr]
{\ln}\frac{|\theta'/\pi|}{2 - |\theta'/\pi|} .
\label{RG}
\ee
These results can be systematically extended to include the regime of 
finite $\sigma'$~\cite{pbv}.
An interpolation between the well known weak
coupling results and those of Eq. (\ref{RG}) is sketched in Fig. \ref{Fig2}.

\noindent{$\bullet $}{\em Conductance fluctuations.~~~}
Rather than expanding $S'_{bulk}$ in powers of $C[q]$ we next elaborate on 
the leading order result, valid for $0 < \theta (\nu ) <\pi$ 
\begin{eqnarray}
S'_{bulk} [q] & = & F(\theta ) - \ln \int d\theta'  P(\theta' ) e^{i\theta' C[q]} .
 \label{P} \\
P(\theta' ) & = & (1-\frac{<\theta'>}{2\pi}) \delta(\theta' ) 
+ \frac{<\theta'>}{2\pi} 
\delta (\theta' -2\pi) .\nonumber
\end{eqnarray}
Here, $P(\theta' )$ indicates that the quantity $\theta'$ is actually
distributed in an highly non-gaussian manner.
The expectation $<\theta'>$ is precisely the same as $\theta'$ in Eq. (\ref{Theta}).
Notice that in the `quantum Hall phase' $<\theta'>$ vanishes exponentially 
in $L^2$, along with the higher order moments. 
At the transition, however, we have 
$<\theta'> = \sqrt{<(\theta' - <\theta'>)^2 >} = \pi$ 
which is large $N$ version of the {\em broad mesoscopic conductance 
fluctuations} that typically occur at the quantum Hall plateau 
transitions~\cite{cp}.

\noindent{$\bullet $}{\em Twisted boundary conditions.~~~}
Following 't Hooft's idea we next employ Eq. (\ref{P}) to compute the response of 
the {\em bulk} of the system to {\em twisted} boundary conditions~\cite{llp}. 
For this purpose we evaluate $S'_{bulk}$
with $C[q] = \frac{1}{2}$ relative to the result with integral topological charge
\be
\Delta S'_{bulk} = - \ln(1-<\theta'>/\pi ) .
\ee
Notice that for $\theta(\nu ) < \pi$ the response is exponentially small in 
the area of the system, $\Delta S'_{bulk} \approx <\theta'>/\pi  
\approx e^{-L^2 /\xi^2 (\nu )}$, 
as expected. However, as one approaches the transition, 
$\theta(\nu) = \theta' = \pi$, 
the response diverges logarithmically. Hence, {\em gapless} bulk excitations 
must exist at $\theta (\nu ) =\pi $. This is in spite of the fact the 
large $N$ system is {\em insensitive} to a change from {\em periodic} to 
{\em anti-periodic} boundary conditions.
\vskip 1mm
The research was supported in part by FOM and INTAS 
(Grant 99-1070).

\end{multicols}
\end{document}